\documentclass[twoside,11pt]{article}

\usepackage{doc_style_fixed}
 \usepackage{amsmath}
\heading{}{}{}{}{}{Foulkes, Thaweethai, Scharfstein, Huang, Reeder}
\usepackage{hyperref,endfloat}

\ShortHeadings{Inverse Probability Weighting for Auxiliary Variable Sampling}{Foulkes et al.}
\firstpageno{1}

\begin{document}

\title{Inverse probability weighting for auxiliary variable dependent sampling in observational studies of Long COVID}

\author{
\name Andrea S. Foulkes \email afoulkes@mgh.harvard.edu \\
\addr Biostatistics Center, Massachusetts General Hospital \\
Department of Biostatistics, Harvard Medical School \\
Boston, MA, USA 
\AND
\name Tanayott Thaweethai \\
\addr MGH Biostatistics, Harvard Medical School
\AND
\name Daniel O. Scharfstein \\
\addr University of Utah School of Medicine
\AND
\name Weixing Huang \\
\addr Massachusetts General Hospital
\AND
\name Harrison T. Reeder \\
\addr Massachusetts General Hospital, Harvard Medical School
}

\author{%
\name Andrea S. Foulkes \email afoulkes@mgh.harvard.edu \\
\addr Center for Biostatistics, Massachusetts General Hospital \\
Department of Medicine, Harvard Medical School \\
Department of Biostatistics, Harvard T.H. Chan School of Public Health \\
Boston, MA, USA 02114 
\AND \\
\name Tanayott Thaweethai \\
\addr Center for Biostatistics, Massachusetts General Hospital \\
Department of Medicine, Harvard Medical School  \\
Department of Biostatistics, Harvard T.H. Chan School of Public Health \\
Boston, MA, USA 02114 \\
\AND 
\name Daniel O. Scharfstein \\
\addr Division of Biostatistics, Department of Population Health, University of Utah School of Medicine \\
Salt Lake City, Utah 84108 
\AND \\ 
\name Weixing Huang \\
\addr Center for Biostatistics, Massachusetts General Hospital \\
Boston, MA, USA 02114
\AND \\
\name Harrison T. Reeder \\
\addr Center for Biostatistics, Massachusetts General Hospital \\
Department of Medicine, Harvard Medical School \\
Boston, MA, USA 02114
}

\maketitle

\begin{abstract}%
Selective testing based on values of auxiliary variables is an increasingly popular design strategy in observational studies and is ubiquitous in electronic health record data. Ignoring this underlying sampling mechanism can lead to biased estimation and erroneous scientific conclusions. Yet, rigorous analytic methods for accounting for two-phase sampling designs in observational settings remain under-utilized. Motivated by the Researching COVID to Enhance Recovery (RECOVER) Adult and Pediatric observational cohort studies, we describe common pitfalls and an approach for analysis of data collected via auxiliary variable dependent sampling.  \end{abstract}


\section{Introduction}
Long COVID (LC) is estimated to afflict between 5 and 10\% of individuals after first infection with SARS-CoV-2 \citep{thaweethai_development_2023,geng_2024_2025,gross_characterizing_2024}, and yet its mechanistic underpinnings remain poorly understood. Early evidence suggests it is associated with multi-organ dysfunction, including pathologies in the blood vessels, brain, heart, gastrointestinal tract, immune system, kidneys, liver, lungs, pancreas, reproductive system, and/or spleen \citep{davis_long_2023}, with varying presentations across individuals \citep{thaweethai_development_2023,geng_2024_2025,gross_characterizing_2024}. Hypothesized mechanisms of LC pathogenesis include immune dysregulation, microbiota dysbiosis, autoimmunity and immune priming, blood clotting and endothelial abnormalities, and dysfunctional neurological signaling \citep{davis_long_2023}. 

A primary aim of the Researching COVID to Enhance Recovery (RECOVER) observational cohort studies was to evaluate these potential mechanisms in both adult and pediatric populations. To this end, RECOVER studies leveraged cost-efficient designs involving prioritized testing based on data collected on auxiliary variables. Auxiliary variable dependent sampling has been described for epidemiologic studies using case-cohort and case-control designs \citep{prentice_case-cohort_1986,breslow_statistics_1996,breslow_using_2009}, electronic health records (EHR)-based studies \citep{zhang_patient_2024,amorim_two-phase_2021,levis_double_2024}, and immune correlate analyses of vaccine efficacy trials to measure candidate surrogate endpoints \citep{follmann_augmented_2006,fong_calibration_2015,fu_joint_2017,gilbert_four_2024}. Such designs are generally categorized as two-phase sampling designs, in which inexpensive variables (collected in phase one) are used to inform the measurement of more expensive variables (collected in phase two). Methodologies for addressing the selection bias introduced by such designs have been widely studied \citep{robins_estimation_1994,rotnitzky_semiparametric_1995,breslow_maximum_1997,breslow_large_2003,wang_causal_2009,rose_targeted_2011,kennedy_efficient_2020,hejazi_efficient_2021}.

In the simplest design setting, study participants reach a pre-defined study visit at which time they may be selected to receive an additional assessment based on a randomization schema that is dependent on one or more variables. For example, in resource-efficient cohort study designs an optimal subset of participants may be selected for additional follow-up using information on current health status. In two-phase vaccine trials, an inexpensive auxiliary variable may be measured at a pre-defined study time point and a subset may be selected for measuring a more expensive outcome \citep{gilbert_optimal_2014}. In a similar fashion, participants in the Researching COVID to Enhance Recovery (RECOVER) Pediatric cohort study were selected based on data collected during their first study visit to be included in the longitudinal follow-up phase of the study \citep{gross_researching_2024}.

More complex designs incorporate sequential opportunities to sample participants for additional assessments based on auxiliary variables that vary from visit to visit. For example, participants in the RECOVER Adult cohort study had repeated opportunities to be selected for additional assessments over a sequence of study visits, with sampling probability at each visit based on the current values of auxiliary variables. Sampling was further limited to participants who met specific eligibility requirements and who did not complete the assessment in a prior window of time. Such a design exemplifies general two-phase strategies in which repeated sampling depends on longitudinal eligibility requirements and auxiliary variables, as well as the result of sampling at prior visits. In RECOVER Adult, there were 31 such tests administered based on a sequential auxiliary variable dependent sampling \citep{horwitz_researching_2023}.

In the context of auxiliary variable dependent sampling, many hypotheses focus on questions involving the relationship between an exposure and an outcome of a test or survey where one or both were measured in subsets of participants. For example, in the RECOVER Adult and Pediatric cohort studies, interest is in evaluating the association between Long COVID (LC), as an exposure, and the result of these so-called “tiered” tests. While the study design requires acknowledging the role of auxiliary covariates in the tiered testing scheme, these auxiliary covariates in some cases arise after the exposure is measured or are otherwise not of scientific interest to adjust for when comparing outcomes by group. Therefore, in this manuscript, we describe an inverse-weighted estimator for outcome means and mean differences by LC status in this setting, adjusting for a specified covariate set that may be distinct from the auxiliary variables used for sampling. Specifically, we consider data subject to two-phase sampling strategies in addition to customary challenges in observational cohort studies, including confounding and data missingness. Herein we present methods in settings where up to one outcome measurement is collected per person, both in settings in which there is a single sampling event and those in which study participants undergo repeated opportunities for sampling. RECOVER Adult and Pediatric observational cohort studies are used as illustrative examples.  

\section{Methods}

\subsection{Auxiliary variable dependent (“tiered”) sampling designs}
\begin{figure}
\caption{Auxiliary variable dependent sampling within the RECOVER Adult observational cohort} \label{fig:adult-design}
\centering
\includegraphics[scale=0.7]{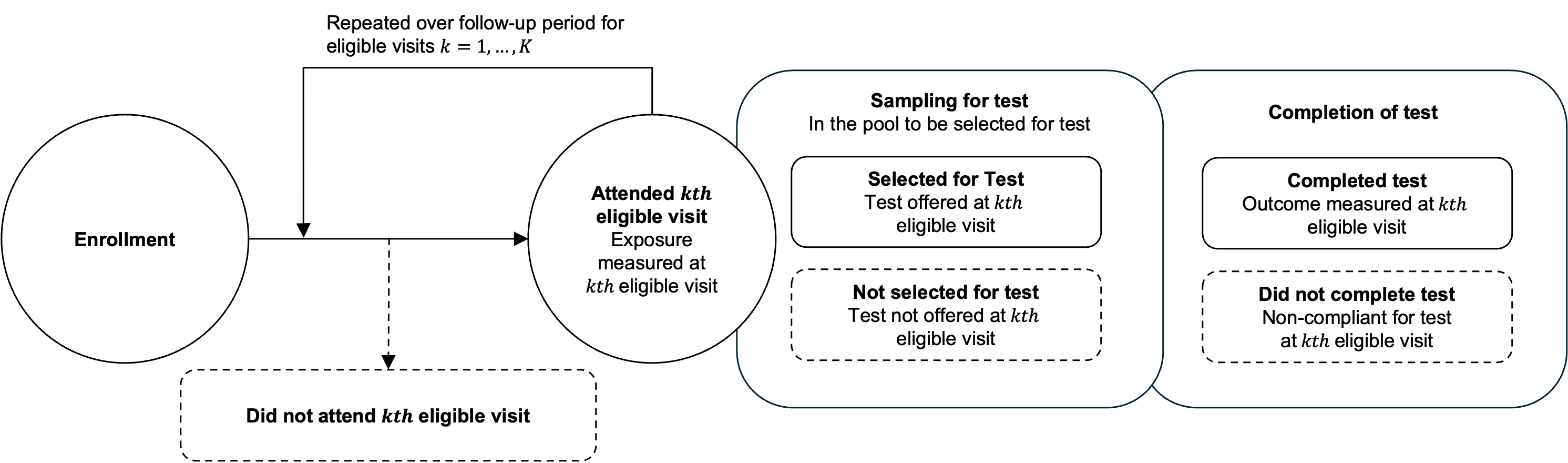}
\end{figure}

Auxiliary variable dependent sampling, also referred to as “tiered” testing, in observational studies can take on many forms. Herein, we consider two distinct designs involving auxiliary variable dependent sampling. In the first case (Figure~\ref{fig:adult-design}), RECOVER-Adult, sampling occurred separately for each of multiple assessments, and with repeated opportunities for selection over multiple study visits. All participants who attended a study visit at which they satisfied the eligibility criteria for a specific assessment were eligible to be sampled for that test, making up the test-specific sampling pool. The exposure was measured in all participants attending an eligible visit. A subset of the individuals in the sampling pool was then selected based on pre-specified probabilities defined by one or more auxiliary variables, and a subset of those selected completed the tier 2 assessment for which they were selected. This auxiliary variable dependent sampling process occurred independently for each assessment and over multiple study visits conditional on triggers, infection status, and eligibility criteria.

\begin{figure}
\caption{Auxiliary variable dependent sampling within the RECOVER Pediatric observational cohort} \label{fig:peds-design}
\centering
\includegraphics[scale=0.7]{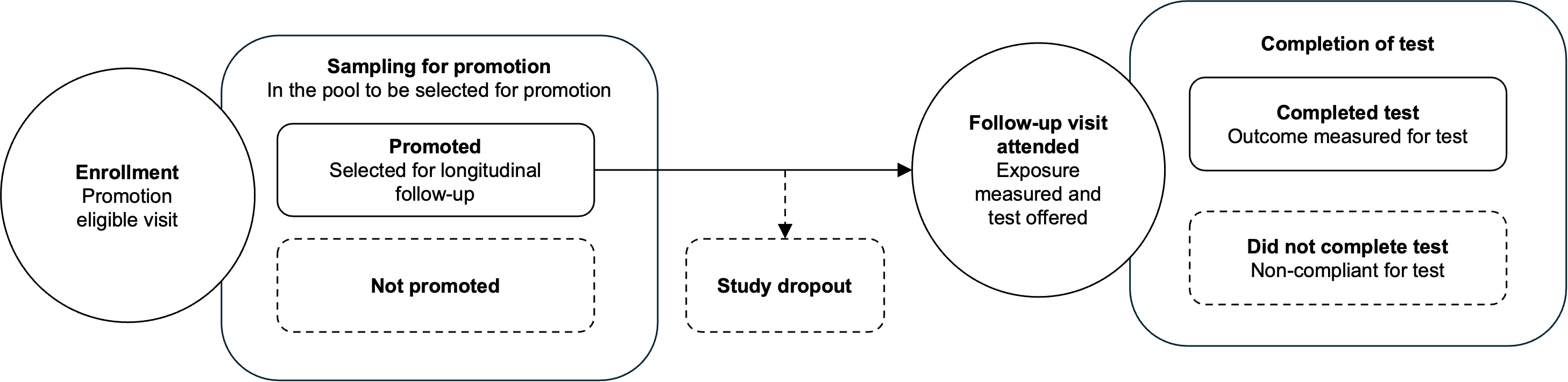}
\end{figure}

In the second case (Figure~\ref{fig:peds-design}), RECOVER-Pediatrics, auxiliary variable dependent sampling occurred only once but prior to measurement of the exposure. In this setting, a subset of the cohort was selected for longitudinal follow-up based on a sampling scheme that depended on auxiliary variables. All participants within this selected group who also attended the longitudinal follow-up visit had their exposure measured and were offered all tier 2 assessments. A subset of participants then completed each tier 2 assessment. A listing of the tiered tests performed in RECOVER Adult and Pediatric cohort studies is provided in Table~\ref{tab:tiered-tests}.  

\begin{table}
\caption{Tiered assessments in RECOVER Adult and Pediatric observational cohort studies} \label{tab:tiered-tests}
\centering
\includegraphics[scale=0.52]{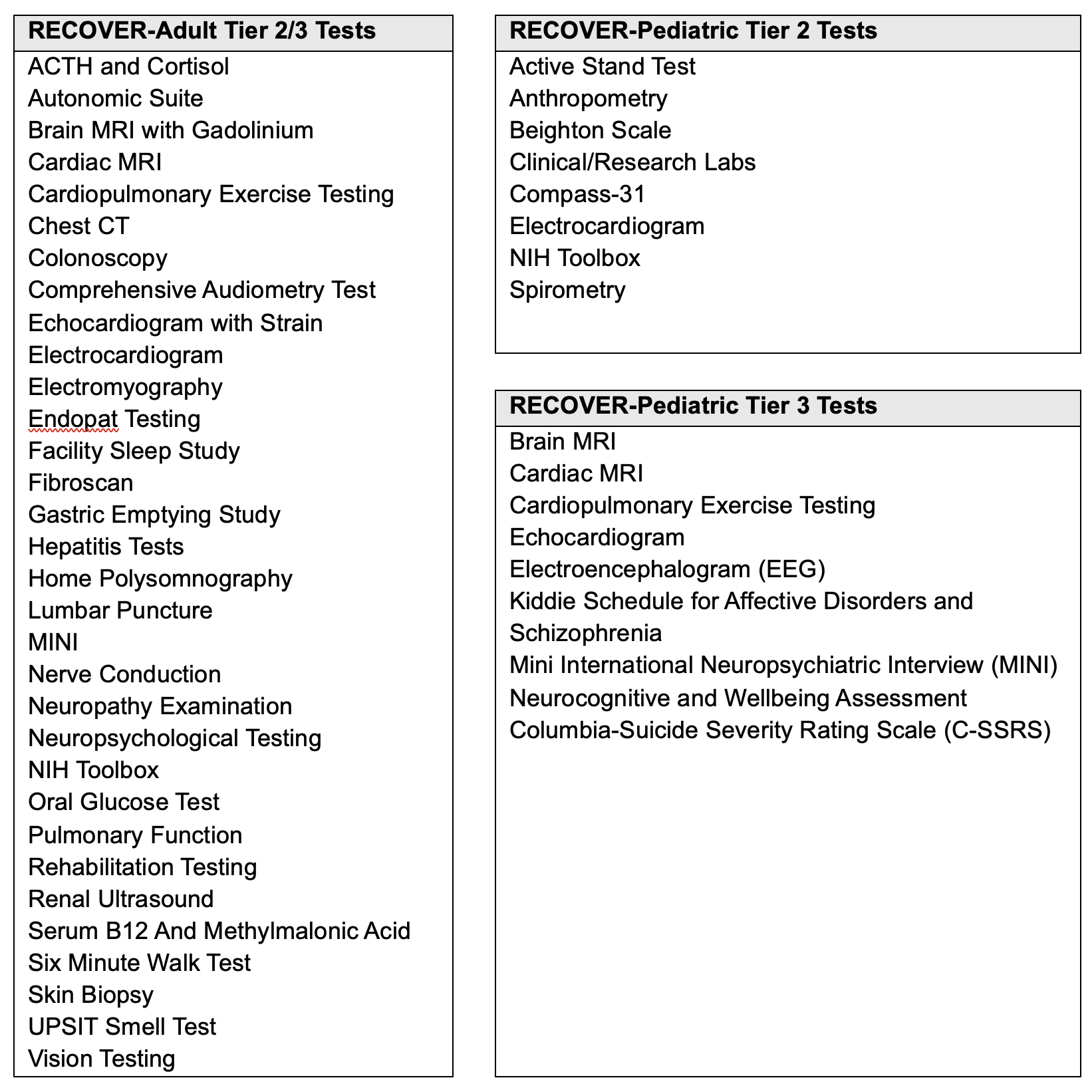}
\end{table}

\subsection{An inverse-weighted estimator in the repeated sampling setting: RECOVER-Adult}

To begin, we consider the RECOVER-Adult design in which participants were subject to repeated opportunities to be sampled for each tiered test. We aim to evaluate the hypothesis that, in a population from which the RECOVER-Adult cohort was drawn, Long COVID (LC) status (the “exposure”) is associated with a concurrent abnormal test (the “outcome”), unconfounded by baseline factors. In this auxiliary variable dependent sampling setting, data analytic complexities include: 1) dropout and other exclusion criteria that render the individuals in the sampling pool at each eligible visit different than the full cohort; 2) imbalances across exposure groups that confound the exposure-outcome relationship; and 3) biased sampling and selective test completion that depends on post-exposure data, resulting in difference between individuals completing tiered tests and the full cohort. 

Herein, we consider a unified analytic strategy that simultaneously addresses loss-to-follow-up during the repeated sampling procedure, the sampling design, differential test completion rates, and covariate imbalance between exposure groups. Formally, for each of a sequence of visits, $k=1,\hdots,K$, and baseline covariates $W$, we define the adjusted means $\mu_k^a=E[E[Y_k\mid A_k=a,W]]$ where $A_k$ denotes exposure status and $Y_k$ denotes the outcome, such that for a binary exposure the adjusted group difference is $\psi_k=\mu_k^1-\mu_k^0$. Aligned with the typical research questions of interest, our estimands of interest are the pooled means of the outcome for each exposure group, and the pooled mean difference between exposure groups, where pooling is over the $k$ visits. To estimate these quantities, we propose an inverse-weighted estimation framework that combines weights based on several probabilities.

At visit $k$, let $U_k$ be an indicator for attending and meeting eligibility requirements for testing and $R_k$ be an indicator for being sampled and completing the test. To facilitate analysis, we consider a monotone missingness structure, such that $U_k=0$ implies $U_{k+1}=0$. In addition, we consider at most one outcome per person such that if $U_k=1$  and $R_k=1$ then $U_{k+1}=0$. For all variables, let an overbar denote the observed history up to and including the specified visit, e.g., the history of test completion up to and including visit $k$ is denoted $\overline{R}_k$. Let $Z_k$ represent auxiliary variables (e.g., trigger variables) that affect missingness and tiered test sampling and completion but are not included as adjustment variables. The conditional probability of attending the $k$th visit among those who attended the prior visit but have not yet had the tiered assessment is
\begin{eqnarray} \label{eq:Ucond}
\pi_k^U(\overline{A}_{k-1},\overline{Z}_{k-1},W)=P(U_k=1 \mid  U_{k-1}=1,R_{k-1}=0,\overline{A}_{k-1},\overline{Z}_{k-1},W)    \;\;  \text{for} \;\;  k>1 .
\end{eqnarray}

\noindent As the cohort is defined based on meeting the eligibility criteria for the test for at least one study visit, by definition $U_1=1$ and so $\pi_1^U=1$.

Among eligible participants who attend the $k$th visit and have not had the tiered assessment at a prior visit, denote the probability of sampling and completion at that visit as
\begin{eqnarray}\label{eq:Rcond}
\pi_k^R(\overline{A}_k,\overline{Z}_k,W)=P(R_k=1\mid U_k=1,\overline{R}_{k-1}=0,\overline{A}_k,\overline{Z}_k,W).
\end{eqnarray}

\noindent Based on these probabilities, define the cumulative probability of consistently being in the study and eligible for testing through visit $k$, without yet completing the test, as
\begin{eqnarray}\label{eq:gU}
g_k^U (\overline{A}_{k-1},\overline{Z}_{k-1},W)=\pi_k^U(\overline{A}_{k-1},\overline{Z}_{k-1},W) \prod_{l=1}^{k-1} \pi_l^U(\overline{A}_{l-1},\overline{Z}_{l-1},W) \left[1-\pi_l^R(\overline{A}_{l},\overline{Z}_{l},W)\right],
\end{eqnarray}

\noindent and the corresponding probability of having the tiered test first performed at visit k as
\begin{eqnarray}\label{eq:gR}
g_k^R (\overline{A}_k,\overline{Z}_k,W)=\pi_k^R(\overline{A}_{k},\overline{Z}_{k},W) g_k^U (\overline{A}_{k-1},\overline{Z}_{k-1},W).
\end{eqnarray}

\noindent Finally, to account for imbalance between individuals with and without the exposure in terms of baseline covariates, define the conditional probability that LC status is equal to $a$ at visit $k$ by
\begin{eqnarray}\label{eq:piA}
\pi_k^A(a;W)=P(A_k=a\mid W).
\end{eqnarray}

\noindent Note that this is the propensity score is defined with respect to the entire study cohort, without further conditioning on those who remain on-study and eligible for their first tiered test at visit $k$.

Using estimates of the probabilities defined in equations (4) and (5), an inverse probability weighted estimator for the adjusted mean $\mu_k^a$ for exposure level $A_k=a$ at visit $k$ can be derived as
\begin{eqnarray}
\widehat{\mu}_k^a=\left[\sum_{i=1}^n \frac{U_{k,i} R_{k,i} I(\overline{R}_{k-1,i}=0)}{\widehat{g}_k^R (\overline{A}_{k,i},\overline{Z}_{k,i},W_i)} \frac{\mathbb{I}(A_{k,i}=a)}{\widehat{\pi}_k^A (a;W_i )} \right]^{-1} \sum_{i=1}^n \frac{U_{k,i} R_{k,i} \mathbb{I}(\overline{R}_{k-1,i}=0)}{\widehat{g}_k^R (\overline{A}_{k,i},\overline{Z}_{k,i},W_i ) } \frac{ \mathbb{I}(A_{k,i}=a)}{\widehat{\pi}_k^A (a;W_i)} Y_{k,i}.
\end{eqnarray}

\noindent The corresponding estimator for the difference in means is given by $\widehat{\psi}_k=\widehat{\mu}_k^1-\widehat{\mu}_k^0$.

This estimator can be implemented by fitting regression models for the probability of attendance~\eqref{eq:Ucond}, completion~\eqref{eq:Rcond}, and propensity score~\eqref{eq:piA} for each visit $k=1,\hdots,K$, with choices guided by sample size and desired model complexity. For example, separate logistic regression models for~\eqref{eq:Ucond} and~\eqref{eq:piA} can be fit at each eligible visit, while a pooled logistic regression model for the visit of first completion can be fit across all visits to estimate~\eqref{eq:Rcond}. Importantly, because the propensity score~\eqref{eq:piA} is defined for the whole cohort but can only be fit on participants on-study and eligible for testing at visit $k$, models used to estimate the propensity score must be inverse weighted by the probability defined by equation~\eqref{eq:gU}. This step is necessary because the propensity score depends on $W$, but does not depend on other auxiliary variables that affect dropout, sampling and completion, and therefore additional weighting is necessary to capture this dependent missingness.

Results are aggregated over all eligible visits using an inverse variance weighted mean, e.g., the pooled mean difference
\begin{eqnarray}
\widehat{\psi}=\frac{\sum_{l=1}^K \widehat{\text{Var}}(\widehat{\psi}_l)^{-1}\widehat{\psi}_l}{\sum_{l=1}^K \widehat{\text{Var}}(\widehat{\psi}_l )^{-1}}.
\end{eqnarray}

\begin{figure}
\caption{Step-by-step approach to analyzing tiered test data: RECOVER Adult} \label{fig:step-adult}
\centering
\includegraphics[scale=0.8]{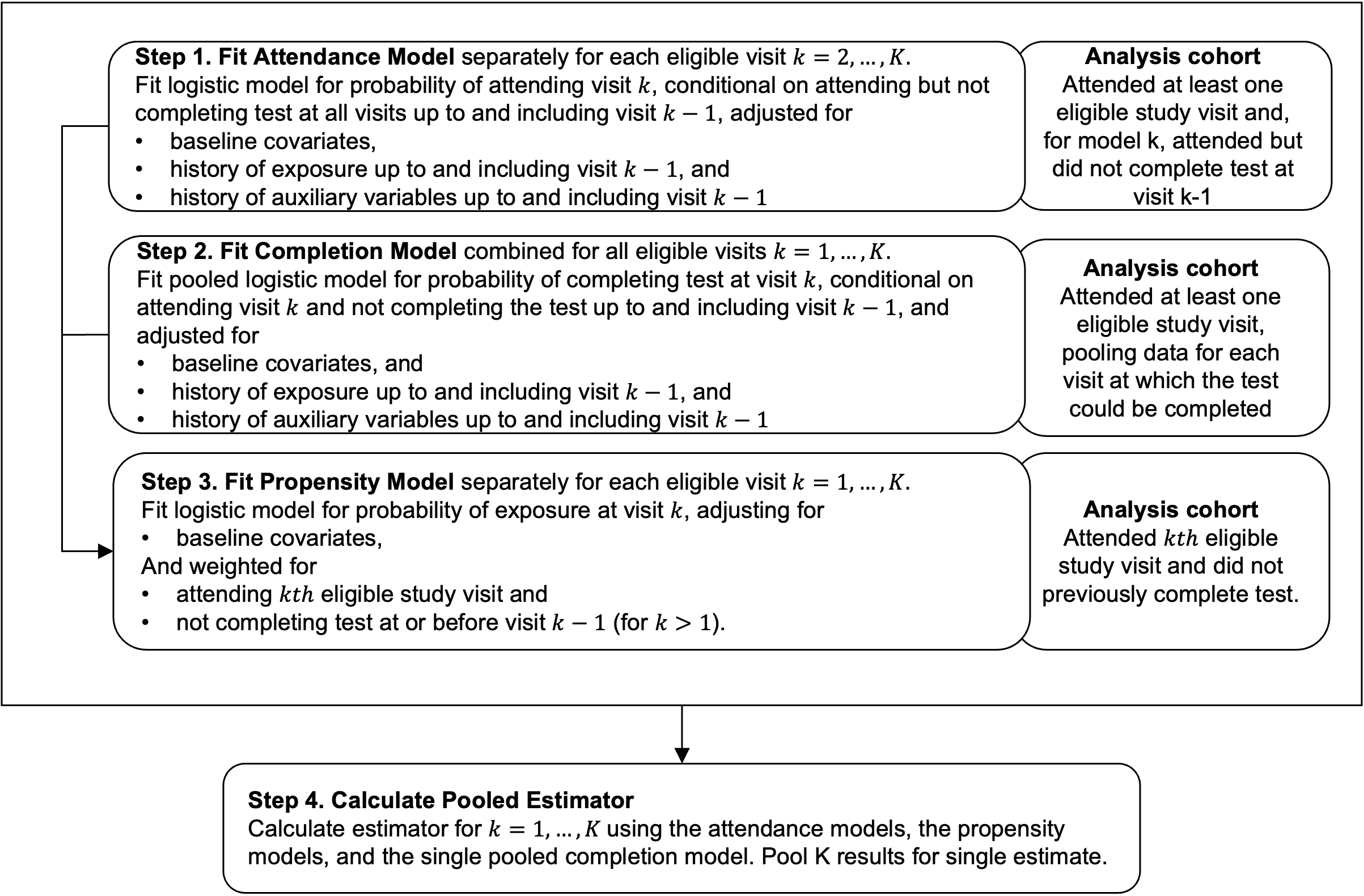}
\end{figure}

A step-by-step summary of this approach is provided in Figure~\ref{fig:step-adult}. Finally, to estimate the variance of each $\widehat{\mu}_k^a$ and $\widehat{\psi}_k$ one can conservatively use standard sandwich variance formulae for M-estimators \citep{vaart_asymptotic_1998}, or estimate the variance via bootstrapping \citep{efron_introduction_1994}, and subsequently estimate variance for the pooled estimator using formulae for the variance of a weighted mean.

\subsection{A modified estimator when exposure is measured after sampling: RECOVER-Pediatric}

In the context of RECOVER-Pediatric, again the estimands of interest are the adjusted means and mean differences of an outcome between exposure groups. In this setting, however, both are measured at a visit subject to auxiliary variable dependent sampling, and each are also subject to missingness. Formally, we define the adjusted mean outcome $\mu^a=E[E[Y\mid A=a,W]]$ and group difference $\psi=\mu^1-\mu^0$. Again, we consider an inverse probability weighted estimation approach to address these challenges, defined in terms of a sequence of selection and exposure probabilities.
In RECOVER-Pediatric, the initial cohort of participants have baseline covariates $W$ and a set of baseline auxiliary variables $Z$.  Based on $Z$, a subset of participants were selected or ``promoted", denoted by the indicator $R$, to attend an additional visit referred to as a Tier 2 visit, with selection probability 
\begin{eqnarray} \label{eq:piRpeds}
\pi^R(Z,W)=P(R=1\mid Z,W).
\end{eqnarray}

Among selected participants, attendance at the follow-up study visit is denoted by the indicator $U$, the occurrence of which may depend on $W$ and $Z$, with probability
\begin{eqnarray} \label{eq:piUpeds}
\pi^U(Z,W)=P(U=1\mid R=1,Z,W).
\end{eqnarray}

Participants who were selected and attend the Tier 2 visit had their LC status A assessed, for which we define a propensity score conditional on the baseline covariates W. This probability is defined with respect to the entire cohort unconditionally, such that the probability that LC status is equal to a at the Tier 2 visit is
\begin{eqnarray} \label{eq:piApeds}
\pi^A(a,W)=P(A=a \mid  W).
\end{eqnarray}

Finally, among those who attend the Tier 2 visit, the completion of the outcome assessment is denoted C, and is subject to missingness that again may depend on both W and Z, as well as A. This conditional probability of completing the Tier 2 outcome assessment is given by
\begin{eqnarray} \label{eq:piCpeds}
\pi^C(A,Z,W)=P(C=1 \mid  U=1,R=1,A,Z,W) .
\end{eqnarray}

Using estimates of each probability defined in equations (6-9), an inverse probability weighted estimator for the adjusted mean for exposure level $A=a$ can be derived as

\begin{equation} \label{eq:ipwpeds}
\begin{aligned}
\widehat{\mu}^a = & \left[\sum_{i=1}^n \frac{R_i U_i C_i}{\widehat{\pi}^R(Z_i,W_i ) \widehat{\pi}^U (Z_i,W_i ) \widehat{\pi}^C(A_i,Z_i,W_i )} \frac{I(A_i=a)}{\widehat{\pi}^A(a;W_i)} \right]^{-1} \\
& \times \sum_{i=1}^n \frac{R_i U_i C_i}{\widehat{\pi}^R (Z_i,W_i) \widehat{\pi}^U(Z_i,W_i) \widehat{\pi}^C (A_i,Z_i,W_i)}\frac{I(A_i=a)}{\widehat{\pi}^A(a;W_i)}Y_i.
\end{aligned}
\end{equation}

\begin{figure}
\caption{Step-by-step approach to analyzing tiered test data: RECOVER Pediatrics} \label{fig:step-peds}
\centering
\includegraphics[scale=0.8]{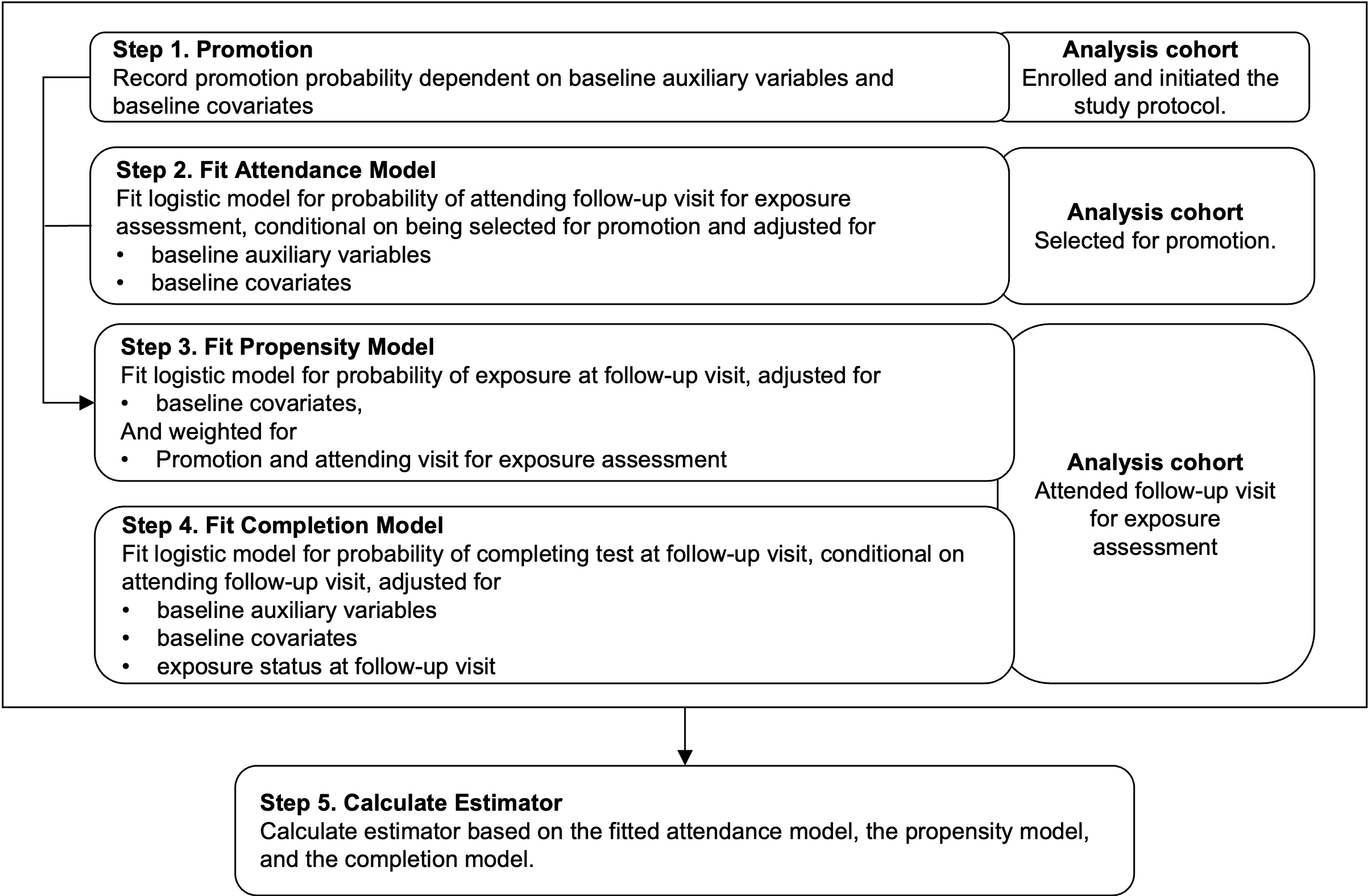}
\end{figure}

In RECOVER-Pediatric, the sampling probability~\eqref{eq:piRpeds} is known by design and does not need to be estimated. The attendance probability~\eqref{eq:piUpeds} and exposure probability~\eqref{eq:piApeds} can be estimated based on sequential logistic regression models fit among the promoted cohort ($R=1$) and the attending cohort ($U=1$), respectively. Importantly, the propensity model for LC status $A$ is also fit among the attending cohort, but must be inverse weighted by the probability of promotion and attendance $g^{R,U}(Z,W)=\pi^R (Z,W) \pi^U(Z,W)$ to correctly reflect the probability of LC in the full cohort. Again, this is necessary because while attendance depends on auxiliary variables $Z$, the propensity score does not, and therefore this dependence must be accounted for by weighting. Finally, the completion probability~\eqref{eq:piCpeds} can also be estimated via a logistic regression model among the attending cohort ($U=1$). Estimation of $\widehat{\phi}=\widehat{\mu}^1-\widehat{\mu}^0$ follows directly, and statistical inference is analogous to the RECOVER-Adult case described above. A step-by-step summary of this approach is provided in Figure~\ref{fig:step-peds}.

\section{Example: RECOVER-Adult}
\emph{Background}. RECOVER-Adult is an observational cohort study of 14,779 individuals at least 18 years of age at enrollment, including participants with and without a history of SARs-Cov-2 infection from 33 states and territories in the United States \citep{horwitz_researching_2023}. Enrollment occurred between October 2021 and January 2024, and follow-up of the first cycle ended October 2025. RECOVER-Adult is composed of two sub-cohorts: the Acute Cohort, including individuals enrolled within 30 days after their index date (defined as first SARS-CoV-2 infection for infected participants and any negative test date for uninfected participants), and the Post-Acute Cohort, including individuals enrolled more than 30 days and up to 3 years after their index date. All RECOVER-Adult participants completed extensive surveys every three months and followed a regular cadence of office visits involving clinical evaluations as well as sample collections \citep{horwitz_researching_2023}.

RECOVER-Adult included the administration of 32 tests subject to sampling (Table~\ref{tab:tiered-tests}). For each test, individuals underwent an auxiliary-dependent sampling procedure repeatedly over multiple visits, with selection probability dependent on current infection status and the presence of a test-specific “trigger” comprised of multiple auxiliary variables. In general, a greater likelihood of being offered a test was associated with a poorer outcome. For example, participants with a modified Medical Research Council (mMRC) dyspnea score of 2 or greater, persistent cough, fatigue, post-exertional malaise, or hypoxia at a given study visit – attributes specifically selected because they tend to associate with poor outcomes on pulmonary function tests (PFTs) and chest computed tomography (CT) – were more likely to receive PFTs and a chest CT.

For most tests, if a participant was eligible at a given visit, their probability of being offered the test was 100\% if they were infected and had the test trigger, 5.6\% if they were infected and did not have the test trigger at the visit, and 11.4\% if they were uninfected. Exclusion criteria for eligibility at a particular visit varied by test and included factors such as recent infection, pregnancy, breastfeeding, and being within a year of previously completing the same test. As participants entered the study at varying times after infection, the first visit at which someone was eligible to receive the tiered test occurred at varying times since infection. Sampling was independent across visits and tiered tests, conditional on triggers, infection status, and eligibility criteria. 

\emph{Example}. For illustration, we consider the University of Pennsylvania Smell Identification Test (UPSIT). The trigger for this tier 2 test was self-reported “loss of or change in smell or taste”. The primary hypothesis we consider is that LC status is associated with concurrent severe microsmia or anosmia among adults with a history of suspected or probable SARS-CoV-2 infection. LC status was measured using the 2024 LC Research Index (LCRI) \citep{geng_2024_2025}, as a binary indicator of LCRI$\ge 11$ and the presence of severe microsmia or anosmia was defined as $\le 25$ on UPSIT. The LCRI and UPSIT were measured concurrently. 

All models were adjusted for the following variables measured with respect to first SARS-CoV-2 infection: age, sex assigned at birth, race/ethnicity, era/cohort, hospitalization, vaccination status, comorbidities (obesity, CVD and diabetes), and social determinants of health (indicators for financial hardship, food insecurity, experiences of medical discrimination, skipped medical care due to cost, lack of social support, and less than college education). The attendance model additionally adjusted for LC status at prior visit. The pooled completion model additionally accounted for the test trigger, LC status at the corresponding visit, and time from first infection.

In analyses without weighting or adjustment, among the 966 LC participants who completed UPSIT, 233 (24.1\%) had severe microsmia or anosmia, and among the 1,356 non-LC participants who completed UPSIT, 175 (12.9\%) had severe microsmia or anosmia. After accounting for attendance, sampling, and completion, the estimated frequences of severe microsmia or anosmia, were 15.3\% and 8.1\%, in the LC and non-LC groups, respectively (Table~\ref{tab:example}). In a fully weighted model accounting for attendance, sampling, and completion as well as imbalance between the LCRI groups, the estimated frequences were 14.6\% and 8.0\%, corresponding to an estimated difference of 6.3\% (95\% CI: 1.3\%, 11.3\%).

\begin{table}
\caption{Long COVID and severe microsmia or anosmia in RECOVER-Adult} \label{tab:example}
\centering
\includegraphics[scale=0.5]{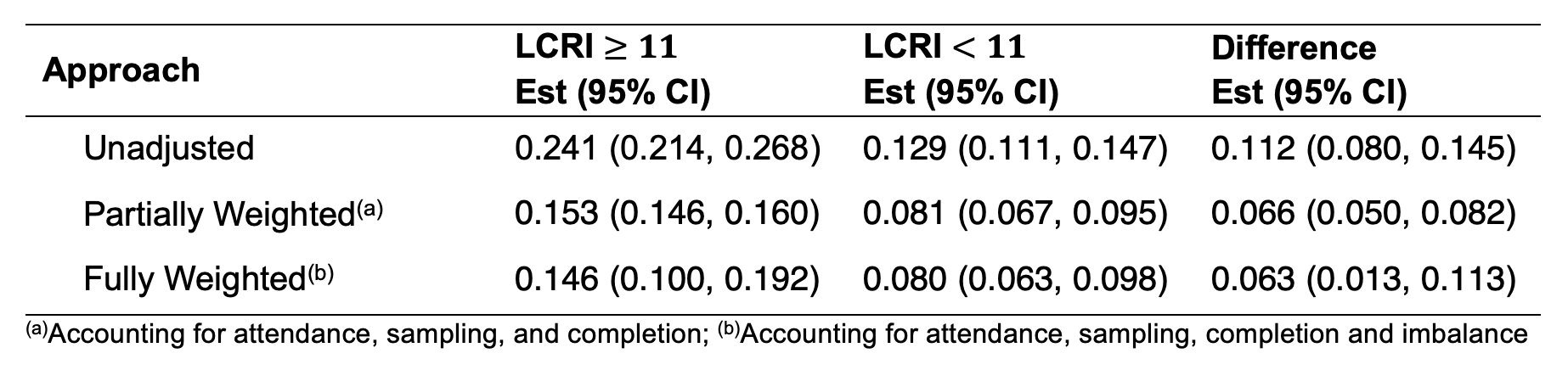}
\end{table}

The reduction in estimated frequencies is consistent with accounting for a sampling schema that favors testing individuals with self-reported loss of smell or taste, a symptom that has been associated with higher rates of severe microsmia and anosmia in SARS-CoV-2 infected individuals \citep{horwitz_olfactory_2025}.  The rates of self-reported change in smell or taste at the first RECOVER study visit at least 6 months after infection were previously estimated to be 40\% among those with LC and 4\% in those without LC \citep{geng_2024_2025}; however, by design, 849 (87.9\%) of participants with LC and 361 (26.6\%) of participants without LC self-reported change in smell or taste at the time of UPSIT completion. As expected, accounting for this oversampling of individuals with self-reported changes who are offered UPSIT reduces the estimated frequency of an abnormal UPSIT finding.   

\section{Discussion}

We presented a unified strategy for analysis of data collected based on auxiliary variable dependent sampling in an observational study setting. Motivated by the RECOVER Adult and Pediatric Observational Cohort study designs, two scenarios were considered. While repeated auxiliary variable dependent sampling was explicitly incorporated in the design of RECOVER, it also occurs implicitly in many observational data settings, including EHR-based studies. An individual captured via EHR has a sequence of encounters, with a wide array of possible assessments conducted at each encounter. Whether a clinician orders a particular test at a given encounter is generally dependent on auxiliary variables measured at that time, such as the presence of symptoms or laboratory results that prompt the need for more testing. Moreover, tests may be performed repeatedly over time. Despite the absence of a formal sampling schema, the resulting data are subject to the same complexities as in studies like RECOVER-Adult: assessments are done at varying times during the overall available follow-up and occur in a biased sample skewing towards those meeting eligibility criteria and test-specific triggers at the time of assessment. An important consequence of sampling a large percentage of individuals with a trigger variable is that those sampled at visits beyond the first eligible visit will disproportionately be those without the trigger variable, i.e., people who never previously triggered the tiered test, and instead were selected based on a sequence of chances to be sampled at random.

The proposed approach focused on hypotheses involving a concurrently measured exposure and outcome, and is a first step towards addressing more general questions. Interest may also be in 1) understanding how an exposure impacts the outcome of a tiered test at a later point in time, 2) characterizing the association between a tiered test finding and another tiered test result, or 3) describing the longitudinal trajectories of repeated tiered testing. Further complexity also emerges if auxiliary variables themselves depend on responses to other tiered tests. For example, the trigger for brain MRI depended on evaluation of UPSIT, itself a tiered test. Likewise, the trigger for Fibroscan depended on the evaluation of laboratory measures that were, apart from enrollment visits, collected only on participants with abnormal findings at earlier study visits if more than six months after first infection. These settings require further consideration due to the potential for additional selection bias.    

In a study for which exposure is defined at baseline, and repeated auxiliary variable dependent sampling occurs subsequently during follow-up, the timing between exposure and the assessment is variable, and dependent on the sampling process itself. Therefore, it should be explicitly acknowledged in this setting that the pooled exposure-assessment association represents a time-averaged relationship over the follow-up period. Further advancements, including doubly robust extensions, as well as approaches to address potential loss of data due to monotonization, are expected to further improve efficiency and robustness.

Repeated auxiliary variable dependent sampling coupled with customary epidemiologic hurdles render the analysis of tiered testing data challenging in observational data settings. Given that the sampling design may favor administration of tests to participants who have abnormal testing findings, accounting for the sampling process is crucial for accurate estimation. Rigorous analytic strategies, as described in this manuscript, that draw on and extend the existing literature on methods for two-phase sampling designs are necessary to facilitate proper inference and, ultimately, help provide greater insight into the mechanisms of disease. These and related challenges resulting from highly complex, dependent sampling schemes and missing data patterns are likely to continue to arise frequently in large-scale observational cohort studies; thus, we aim to raise awareness of the types of estimation strategies and corrections that, when properly combined, can help to rigorously extract reliable evidence from such epidemiologic studies.    

\acks{Support for this research was provided by NIH/NHLBI R01 HL162373 and OT2 HL161841.}

\vskip 0.2in
\bibliography{lc_refs_2026-06-02}

\end{document}